# Exploring Public Perceptions of Generative AI in Libraries: A Social Media Analysis of X Discussions

**Li, Yuan** University of Alabama, USA | yuan.li@ua.edu
**Mandaloju, Teja** University of North Texas, USA | tejababumandaloju@my.unt.edu
**Chen, Haihua** University of North Texas, USA | haihua.chen@unt.edu

## ABSTRACT

This study investigates public perceptions of generative artificial intelligence (GenAI) in libraries through a large-scale analysis of posts on X (formerly Twitter). Using a mixed-method approach that combines temporal trend analysis, sentiment classification, and social network analysis, this paper explores how public discourse around GenAI and libraries has evolved over time, the emotional tones that dominate the conversation, and the key users or organizations driving engagement. The findings reveal that discussions are predominantly negative in tone, with surges linked to concerns about ethics and intellectual property. Furthermore, social network analysis identifies both institutional authority and individual bridge users who facilitate cross-domain engagement. The results in this paper contribute to the growing body of literature on GenAI in the library and GLAM (Galleries, Libraries, Archives, and Museums) sectors and offer a real-time, public-facing perspective on the emerging opportunities and concerns GenAI presents.

## INTRODUCTION

Beginning with the release of ChatGPT in 2022 and followed by tools like Google Gemini, Microsoft Copilot, Claude, and various specialized generative agents, generative AI (GenAI) has seen rapid public adoption and integration across various areas ranging from personal routine uses to professional environments/domains. In Library and Information Science (LIS), both researchers and practitioners are increasingly exploring how GenAI can support libraries to work better regarding organizational processes, enhance user services, transform service delivery, etc.

GenAI tools offer new possibilities for automating routine tasks, streamlining workflows, improving accessibility, and fostering more engaging user experiences. Recent experimental studies have explored GenAI's role in areas such as collection management, subject analysis, and metadata creation (Chen, 2023; Cox & Tzoc, 2023). Studies illustrate GenAI's potential to optimize resources and support users more efficiently (Panda & Kaur, 2023). However, these advancements are accompanied by growing concerns related to data privacy, bias, and transparency, particularly regarding how AI systems make decisions, whose values they encode, and how their outputs influence user understanding (Oddone et al., 2024; Meakin, 2024). In academic contexts, GenAI raises further questions about authorship, critical thinking, and information literacy (Willet & Na, 2024). These tensions emphasize the need for a responsible and context-aware integration of GenAI into LIS practices that ideally involve collaboration among librarians, technologists, educators, and policymakers (Mannheimer et al., 2024).

In parallel with scholarly debates, social media platforms such as X (formerly Twitter), Reddit, and YouTube have become central sites for public discourse on the implications of GenAI. Existing studies highlight that public sentiment toward GenAI is often shaped by users' professional affiliations, media consumption patterns, and personal experiences. For example, Murayama et al. (2025) found more favorable perceptions of image-generating tools like DALL-E compared to text-based models like ChatGPT, with significant variation across language communities. Miyazaki et al. (2024) reported that individuals in tech and creative industries were more enthusiastic about GenAI than those in education or public policy. Qi et al. (2024) observed that while technologically focused communities tend to display polarized views, many others express cautious optimism tempered by ethical concerns.

Despite growing public engagement with GenAI, limited research has focused specifically on how these technologies are discussed in relation to libraries and cultural heritage institutions. Understanding public perceptions in this context is essential, as libraries are not only information service providers but also trusted public institutions that help shape collective knowledge and values. Analysis of the public perception of GenAI in libraries based on social media data can offer valuable insights into how the broader public is imagining the future of these institutions amid rapid technological change.

This study addresses this gap by analyzing public discussions about GenAI and libraries on X. Through content and network analysis, we examine how users perceive GenAI's influence on libraries and identify emerging trends over time. Specifically, we post three research questions:

**RQ1:** How has the volume of posts related to libraries and GenAI evolved over time, and what temporal patterns or peaks can be observed?

**RQ2:** What is the overall distribution of sentiment (positive, neutral, negative) in posts concerning libraries and GenAI and how does sentiment vary across time?

**RQ3:** What does the social network structure reveal about key actors, communities, and patterns of interaction within X conversations on libraries and GenAI?

By identifying the changing trends, sentiment dynamics and social network connections, our findings aim to provide a grounded understanding of public perspectives on automation, ethics, and the evolving role of libraries and ultimately inform more responsible and context-aware GenAI adoption within the field of Library and Information Science.

## RELATED WORK

GenAI presents the library sector with new opportunities to automate routine tasks and enhance the user experiences. Current research in this area focuses on leveraging GenAI to optimize information access, streamline workflows for managing collections, enhance research operations, and improve overall user satisfaction. Meanwhile, studies also examine the perceptions associated with GEnAI adoption related to ethical concerns, legal issues, and educational challenges in the pedagogical context.

Similar to the earlier research in artificial intelligence (AI), recent studies suggest that GenAI provides opportunities for libraries to enhance the effectiveness and efficiency of library workflows and services. These include automating routine tasks such as cataloging, metadata tagging, reference services, research support, teaching, writing and creation, personalization, accessibility, data collection and analysis, and user engagements (Chen, 2023; Cox and Tzoc, 2023; Panda and Kaur, 2023). Lund and Wang (2023) explored possibilities of the implication of ChatGPT for libraries through a dialogue interaction with the tool. Areas they identified include search and discovery assistance, reference and information services, cataloging, and metadata generation. Harada (2025) explored the use of GenAI in automating the assignment of classification numbers to books acquired by libraries. Their findings found that the GenAI model can correctly assign classification numbers to about half of the books from the National Diet Library (NDL) with over 70% accuracy. Their study also noticed that the effectiveness of the approach depends on the quality and quantity of the training data. Semeler et al. (2024) investigated the usefulness of OpenAI Codex with data librarians. Their findings highlighted the GenAI tool's efficiency in assisting data librarians with code script creation for web scraping tasks. The findings suggest that GenAI could be a valuable resource to support data librarians when they are dealing with big data challenges on the web and help enhance data librarianship core competencies like hacker skills and digital literacy.

Along with its speedy development and opportunities, researchers have studied perceptions of GenAI with people involved, ranging from librarians, staff, students, and educators. The findings reveal both advantages and disadvantages perceived from various aspects. Adetayo (2023) surveyed 54 university students regarding their perspectives on the utilization and potential benefits of ChatGPT and the disadvantages over traditional reference librarians. Survey results show that students are aware of the convenience, user-friendliness, extensive knowledge, and accessibility, along with limitations such as comprehending emotions, being unable to answer complex questions or providing incorrect answers, and outdated information. Meakin (2024) examines how generative AI influences library resource consumption among university students and its impact on research habits and learning processes. The study finds that students have mixed views: On the one hand, these tools offer efficient ways to summarize information, extract key insights, and assist with academic writing. On the other hand, they raise concerns about critical thinking, originality, and the ethical use of information. Meakin (2024) emphasizes the urgent need for digital literacy programs to help students critically evaluate the outputs of AI technologies. In the same study, Meakin (2024) found that librarian staff face the challenges of adapting to the environment where AI is changing traditional research methods. The author stresses the importance of professional development to equip library staff with the skills needed to guide students in using AI tools effectively. Adegboye et al. (2024) interviewed health science librarians, and the findings highlighted the challenge of building trust in the reliability and accuracy of ChatGPT, emphasizing the need for evaluation, monitoring, and addressing data privacy concerns. Librarians recognized the importance of caution and transparency when delivering ChatGPT-generated results to patrons. Balancing efficiency with accuracy and security emerged as a critical consideration. Wu et al. (2024) also address concerns by students and faculty surrounding GenAI, including its influence on conventional pedagogical practices and the increasing risks it poses to academic integrity.

Among all the discussions, researchers commonly discuss ethical issues, including privacy concerns, intelligence integrity, and biases. Choi et al. (2023) address potential challenges in its use, including content inaccuracy, lack of consistent system reliability, and ethical considerations, which may impede success in GLAM institutions. Staudt Willet and Na (2024) discussed concerns surrounding GenAI's hindrance to student learning and the potential decline of academic integrity. Suggestions are proposed to address these problems. Oddone et al. (2024) proposed that teacher Librarians are able to draw upon their dual qualifications in education and librarianship to collaborate and add value through a different perspective to the teacher's viewpoint that can lead the TL to engage with GAI tools differently, incorporating them into lessons focused upon the development of specific information literacy skills rather than broader inquiry pedagogies. Meakin (2024) suggests that libraries must remain committed to promoting academic integrity and ethical information practices. Hazarika (2024) and Subaveerapandiyan (2023) propose that collaboration among librarians, AI researchers, and stakeholders is essential for maximizing the potential of generative AI while addressing challenges related to its implementation. Mannheimer et al. (2024) propose adopting ethical frameworks, conducting bias audits, and involving diverse stakeholders in developing and using AI tools. They emphasize the importance of transparency, encouraging libraries and archives to clearly explain the algorithms at work to their users.

The use of generative AI has generated discussions concerning the intent and usage across various areas worldwide. There is burgeoning research addressing perceptions of generative AI via social media platforms. Murayama et al. (2025) analyzed 6.8 million posts on X across 14 languages and found that image-generation tools, like DALL-E, received more positive comments than ChatGPT in most linguistic communities. Miyazaki et al. (2024) analyzed X posts and identified variations in sentiment based on users' occupations and how they use AI. Users in fields like software IT and creative industries tended to be more enthusiastic. In contrast, users in traditional fields like education, law, or policymaking were more likely to express reservations. Staudt, Willet, and Na (2024) examined early reactions to ChatGPT within education-focused subreddits, analyzing the frequency, participation, and content of discussions. Their findings are indicated by multiple roles within the education sector, including educators, students, and stakeholders. The study identifies key themes in the discourse, including the potential of ChatGPT support with educational tasks, individual learning, and a potential decrease in office hours beyond student engagement. Their study also reflected concerns surrounding students and generative AI use, including over-reliance on generative AI and a potential decline in critical thinking skills. Based on the strengths and potential areas of concern, they recommend a thorough review and consideration of generative AI within the existing educational frameworks. Qi et al. (2024) utilized computational and qualitative methods to review and categorize subreddit comments containing keywords related to GenAI, including generalized and specific keywords. The study utilized BERTTopic to analyze 33,912 comments from 388 subreddits between 11/30/2022 and 06/08/2023. Thematic analysis identified recurring topics, including AI consciousness, development, business applications, creativity, and societal impact. The study reflected primarily a positive perception of GenAI especially within decision-making contexts like gaming and education. However, concerns were also expressed regarding some distrust of current AI technology and the potential impact of generative AI on future use. Qi also specified that communities with a heavy technological focus exhibit more polarized sentiments compared to non-technology-focused communities.

In terms of academic research on social media, discussions about GenAI in GLAM institutions are still emerging and relatively limited. Brewer et al. (2024) conducted a survey study with 1035 participants and investigated how media consumption patterns and exposure to specific media messages relate to public attitudes toward AI image generators. They found that people who consume more traditional media are generally more skeptical and view AI-generated art as unethical and even as a form of theft. On the contrary, those who engage more with digital or tech-focused media tend to be more positive and more accepting of AI art as innovative. Kidd et al. (2024) highlight that as museums and galleries adapted their social media strategies to increase their digital engagement during the COVID-19 lockdowns, many users appreciated the continued access to cultural content, while others expressed concerns about digital fatigue and the authenticity of online experiences compared to in-person visits.

Our study addresses this gap by analyzing public opinion on GenAI's role in GLAM institutions through a content analysis of posts on X/Twitter. We aim to capture how people perceive the influence of GenAI in these cultural spaces—offering insights into perceptions, hopes, and concerns from a wider public lens.

## METHODS

In this section, we describe our process of data collection, data preparation, and analysis methods.

### Data collection

We collected post data from X using Sprinklr, a powerful social listening and data aggregation tool that "empowers brands to monitor and analyze conversations across a vast array of 30+ social and digital channels"

(https://www.sprinklr.com/). Our objective was to gather a diverse dataset rich in textual information relevant to the discussion of GenAI and library institutions and services. Given that we want to focus on immediate discussions that reflect the public attention over time, we chose to focus on X. Other platforms like Reddit, Facebook, Instagram, YouTube, and WordPress are not included. To collect data from Sprinklr, we created a search query that includes a list of keywords related to GenAI technology and libraries. The keywords were initially brainstormed based on the authors' expertise and familiarity with the literature, and then refined using preliminary results from a prior study that explored X posts discussing themes related to GenAI and libraries [anon.]. The preliminary study helped us understand the most relevant terms, trends, and variations that would capture a comprehensive dataset. To gain a broader understanding from the public discussion instead of focusing on specific aspects of library institutions and services, we did *not* include any specific terms that could possibly be more familiar to library professionals but not to the public (e.g., metadata, cataloging, reference, help desk, etc.). We also considered irrelevant data that might be caused by different uses of the term "library," especially when it is used to refer to software packages or open-source libraries related to the computer science area.

Based on the identified keywords, we constructed a well-structured search query. This involved logical operators (AND, OR, NOT) to refine the search criteria, ensuring we retrieved high-quality, relevant data while filtering out noise and irrelevant entries. Here is our finalized query used on Sprinklr: *("LLM" OR "GPT" OR "generative pretrained transformer" OR "Generative AI" OR "Chatgpt" OR "LLama" OR "perplexity" OR "Midjourney" OR "CLAUDE" OR "DALL-E" OR "Bard" OR "Copilot" OR "perplexity" OR #GPT OR #LLM OR "ai" OR "GenAI") AND ("library" OR "libraries" OR "librarian" OR #library OR #librarian) NOT ("open source library" OR "python" OR "machine learning" OR #python OR #machinelearning).*

Using Sprinklr's advanced filtering capabilities, we collected English-language data from X over a two-year period, ranging from November 1, 2022 to November 30, 2024. The dataset contains 448,554 posts (rows) and 55 columns that included raw text, metadata (e.g., sender, sender followers count, original authors, timestamps, sources, engagement metrics, geographic information), as well as initial sentiment indicators (i.e., positive, neutral, negative) generated by Sprinklr's built-in sentiment analysis function.

## Data Preparation

We first removed duplicate entries to prepare for data analysis, which reduced the dataset to 448,407 unique rows. We dropped columns that were empty, bringing the total column countdown to 34. This was followed by removing irrelevant or redundant columns, such as "SenderProfileImgUrl" and "SenderScreenName," which do not contribute meaningfully to textual or locational analysis. Columns with minimal or no variance (like those with only one unique value) and those with excessive missing values were also removed to clean and streamline the dataset, for example, location-specific fields. A pair of seemingly duplicate columns, "Message Type" and "MessageType", were found to be identical. To avoid redundancy, one of them was dropped. To prepare the data for meaningful analysis, we also applied text preprocessing to the tweet content. This included cleaning the "Message" field by removing URLs, user mentions, and special characters and converting all text to lowercase. This ensured consistency and removed noise from the textual data, making it suitable for further natural language processing.

Next, to maintain our focus on understanding the intersection between Generative AI and libraries, each tweet was evaluated for relevance using a custom prompt powered by LLaMA 3, a locally hosted large language model via Ollama. The model was then prompted to classify all 112, 046 posts as either "Relevant" or "Irrelevant" based on whether they discussed Generative AI technologies (e.g., GPT, ChatGPT, DALL·E), Libraries, and their services. Mentions of "programming libraries" or unrelated content were automatically marked as Irrelevant. Only the posts classified as Relevant were retained for the final dataset. This significantly refined the corpus, reducing it from 448,407 posts to 215,548, ensuring that the subsequent analysis remains tightly aligned with the research focus.

After isolating posts relevant to Generative AI and libraries, further refinement was applied to ensure language consistency across the dataset. Since the research is focused on English-only posts, posts written in other languages were excluded. A simple function was used to retain only those posts containing strictly ASCII characters, typically representing English text. By applying this filter, the dataset was reduced from 215,548 to 112,046 posts, further sharpening the focus of the analysis. This final set of English-only, relevant posts was then used for deeper explorations, such as keyword and social network analysis, ensuring that linguistic noise or translation ambiguity did not interfere with the insights.

After all the cleaning steps, we have 112,046 relevant posts. Table 1 shows a detailed breakdown based on the post types: *X Updates*–original posts authored by a user, represents new, original content; *X Reposts*– retweet of someone else's tweet, often used to share or endorse content without adding new information; *X Replies*–response to someone else's tweet, which is often a part of conversation; and *X Mentions*–tweet that mentions another user (e.g., @username) for drawing attention or tag people/organizations but not a reply or repost.

**Table 1: Tweet type breakdown by volume, uniqueness, and duplication percentage.**

| Message Type | Total | Unique (Per type) | Duplicates | % of Total | % Unique |
|---|---|---|---|---|---|
| X Update | 12,084 | 11,526 | 558 | 10.78% | 38.10% |
| X Repost | 82,009 | 4,700 | 77,309 | 73.19% | 15.52% |
| X Reply | 15,217 | 11,425 | 3,792 | 13.58% | 37.76% |
| X Mention | 2,736 | 2,649 | 87 | 2.44% | 8.76% |
| **Total** | 112,046 | 30,270 | 81,776 | 100% | 100% |

Out of the 112,046 posts, 30,270 were unique posts, which counts about 27% of the total dataset (note that this is the *true unique count* across the entire dataset as some messages appear in multiple message types but are only counted once in the *true unique count* to avoid double-counting). Within the unique posts, *X Updates* and *X Reply* together account for about 76% of the unique content, even though they comprise only about 24% of the entire dataset. *X Reposts* comprise about 73% of all posts but only contribute 15.5% of unique posts, indicating that most posts are primarily recycled content. We decided to include all the cleaned data (112, 046) for our data analysis to provide a rich, realistic picture of how the public conversations unfold on X about the impact of GenAI on libraries. Meanwhile, we realized potential limitations such as bias toward popular content due to reposts and distorter for qualitative analysis regarding topic trends.

## Data Analysis

*Sentiment Analysis.* As we noticed that there were conflicts between the Sprinklr-generated sentiment tags and the authors' judgment, we decided to recode the sentiment associated with X posts. Research has found that LLMs like LLaMA can be very efficient and reasonably accurate for large-scale sentiment analysis and models like LLaMA exhibit high internal consistency, sometimes surpassing human annotators (Bojic, et al., 2025). We then experimented with LLaMa3 to evaluate the sentiments for a subset of data and identified 200 specific posts where the Sprinklr-generated sentiments did not align with those generated by LLaMa3. These 200 posts were split into two sample sets, and two human coders independently reviewed each set to evaluate the accuracy of the sentiment tags. The agreement calculation shows that, on average, 27% of the human-coded samples aligned with the original Sprinklr-generated sentiments, while 60% aligned with LLaMa3's sentiment labels. Only 13% of the samples were judged as incorrect by both human coders, indicating a higher overall agreement with the LLaMa3-generated sentiments compared to the original labels. This suggests that LLaMa3 provides a more reliable assessment of sentiment in this context. We then used LLaMa3 to recode the sentiment for each of X posts with the following prompt "*You are an experienced sentiment analyst. Your task is to determine the sentiment of a given tweet. Read it carefully and classify the sentiment as 'POSITIVE', 'NEGATIVE', or 'NEUTRAL'. Your response format should be only one of these words: POSITIVE, NEGATIVE, or NEUTRAL.*"

*Social Network analysis.* We conducted social network analysis (SNA) to help understand the dynamics of user interactions within the X dataset. Mentions are a key mechanism on posts through which users engage with others by tagging their usernames using the "@" symbol. Extracting these mentions from each post in our dataset allowed us to build edges between the author of a post and the users they mentioned. This approach provided a way to model the flow of information and identify key actors in the conversation. The mention-based edges were then used to construct a directed graph using NetworkX, where nodes represent individual users and directed edges represent mention relationships. For example, if User A mentions User B in a post, a directed edge is created from A to B. This directed nature of the graph captures the direction of communication and helps identify both the broadcasters and receivers of attention within the network. Three centrality measures were computed to identify the most influential users: degree centrality, betweenness centrality, and eigenvector centrality. Degree centrality highlights users who are most active or most mentioned in the network. Betweenness centrality shows which users act as bridges between different parts of the network, controlling the flow of information. Eigenvector centrality measures the influence of a user based on the influence of their connections, highlighting users connected to other influential nodes.

# RESULTS

For **RQ1**, we focused on the trends of posting numbers over time. Figure 1(a) presents the monthly trend of X posts focusing on GenAI and libraries. There are two top peak months: August 2023, with 9,716 posts, and September 2024, with 46,087 posts. Except for these two months, the average monthly posts are about 2,343. In another words, the posts of August 2023 are about 4 times the average, whereas September 2024 posts are exceptionally high–over 10 times the average. This could be caused by seasonal trends like social or technology events, conferences, product launches, or news coverage that trigger surges in tweet activity.

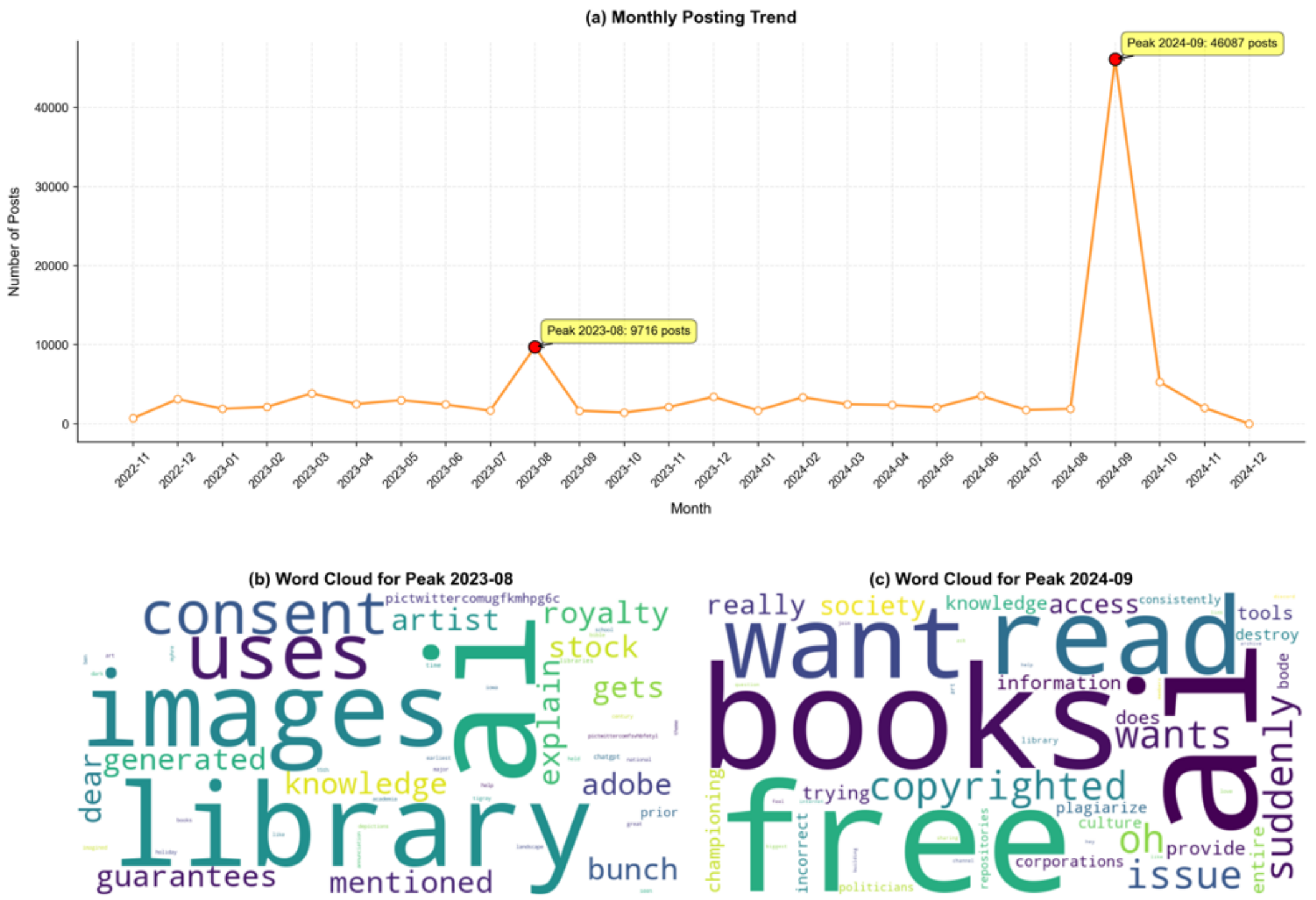

**Figure 1: Monthly posting trends and keyword analysis on libraries & GenAI**. (a) shows monthly distribution of English posts from November 1, 2022 to November 30, 2024. The plot shows the temporal trend in monthly overall post volume. It indicates two distinct peaks in August 2023 and September 2024. (b-c) Word clouds represent the top 50 most frequent keywords during these peak periods, providing insights into the dominant topics of discussion during high-activity months.

Through a keyword analysis of term frequency (Figure 1(b) and Figure 1(c)), we found out that the keywords of August 2023 include *AI, images, library, artist, consent, generated, adobe, knowledge, stock, school, ChatGPT, and royalty*. These words show the focus of AI-generated content, especially images. Meanwhile, *artist, consent, adobe,* and *stock* could imply concerns over intellectual property and ethics, especially for image generation tools. In addition, the presence of *school* may indicate the increased engagement of students during school breaks or summer vacations. Keyword analysis of posts in September 2024 shows that the highly repeated terms are *AI, books, free, read, copyrighted, issue, consistently, plagiarize, championing, politicians,* etc. These keywords imply the discussion focused on copyright, books, and free access to information, which is related to the criticism regarding how GenAI interacts with creative and educational content. Some of these words are probably associated with strong opinions triggered by events like campaigns.

For **RQ2**, we focus on the sentiment presented by the public discussion and the changing trends over time. Figure 2(a)-(d) shows the sentiment trends over time and the peak months where the sentiment count is higher than the surrounding months for each sentiment, respectively Fig. 3(a) to Fig. 3(c) and together for comparison purposes Fig. 3(d). Our analysis of the monthly sentiment trends reveals distinct patterns in the emotional expression for the observed time period. We found that the negative sentiment was the strongest and most volatile, neutral sentiment was more steady, and positive sentiment stayed lower during the entire period (Figure 2(d)).

Particularly, 50, 679 posts are classified as *negative*, accounting for 45.23% of the total posts (Table 2). Eight negative peaks are observed, while the average peak height is about 5,714 per month. The maximum peak observed in September 2024, which includes 35, 078 posts are classified as negative. The month with the minimum peak contains 470 posts as negative in May 2023. The neutral sentiment category includes 39,956 (35.66%) posts with nine peaks with an average of 2,629 posts. The highest peak was observed for September 2024, with 9,425 posts, and the minimum peak was in February 2024, with 1,315 posts. Positive sentiment was the least prevalent compared to the negative and neutral sentiments. With a total of 21,411 posts, eight peaks were observed with an average height of 1,320 posts, the maximum peak observed in October 2024 with 3,176 posts, and the minimum peak in September 2023. Based on these findings, we conclude that the overall tone of the discussion about the impact of GenAI on libraries was predominantly negative, and the neutral sentiment remains more stable. However, the lower frequency and less intense positive peaks suggest the optimistic attitude towards GenAI in libraries was less popular over time.

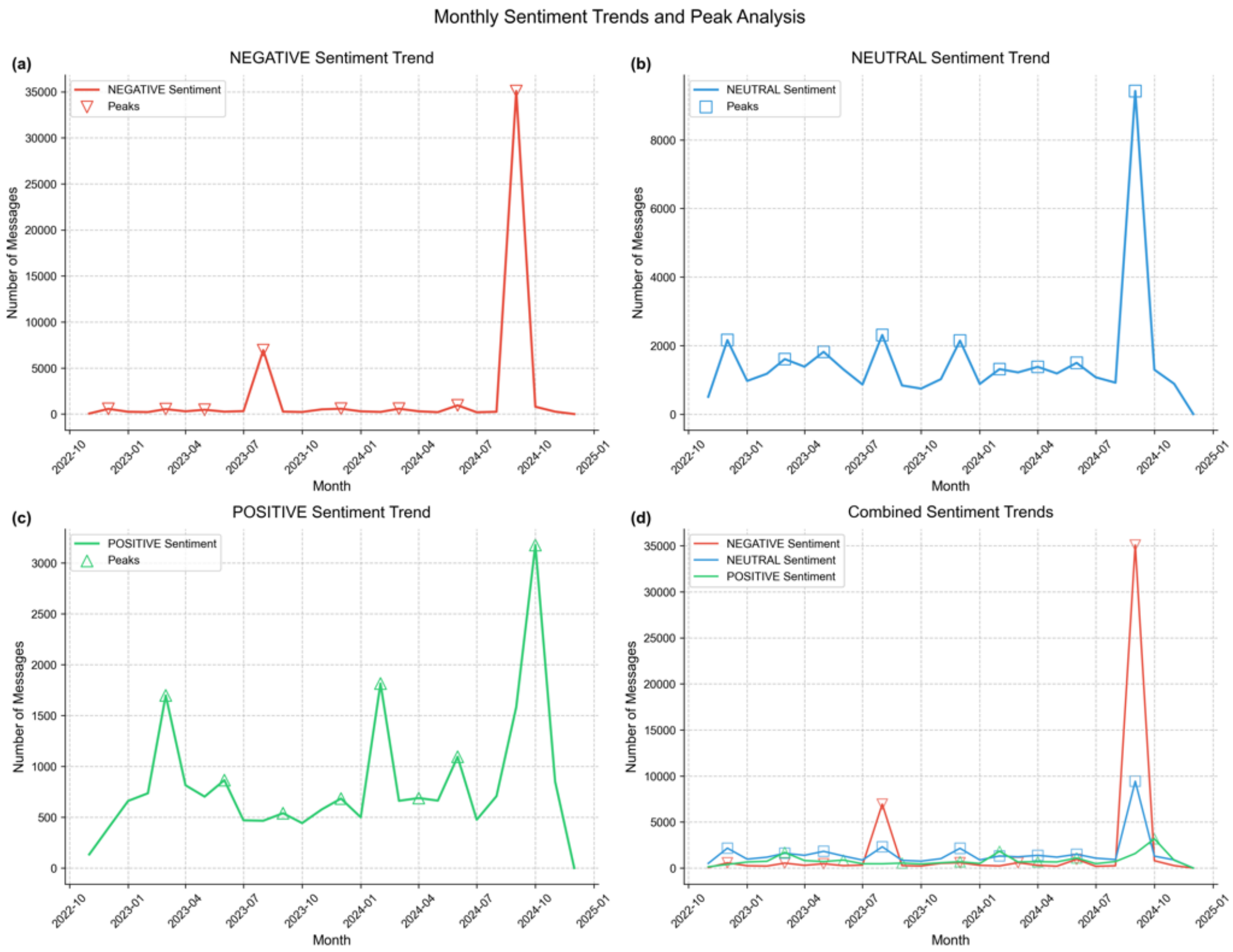

**Figure 2: Monthly sentiment trends and peak analysis from November 1, 2022, to November 30, 2024, on X.** (a) shows the negative sentiment trend with monthly message counts and detected peaks, (b) shows the neutral sentiment trend with monthly message counts and detected peaks, (c) shows the positive sentiment trend with monthly message counts and detected peaks, and (d) shows the combined view of all sentiment trends, illustrating the temporal distribution and relative proportions of different sentiment categories over time. In this figure, peaks are marked with distinct markers (triangles for negative, squares for neutral, and upward triangles for positive sentiments).

**Table 2: Summary of sentiment peaks and distribution across the dataset.** In this table, we summarize sentiment peaks identified in the dataset, including their frequency, intensity, and overall distribution. The data indicates a predominantly negative discourse around GenAI's impact on libraries, with limited yet notable surges in positive sentiment.

| **Sentiment** | **# of Peaks** | **# of Average Peak** | **Median Peak Height** | **Max Peak** | **Min Peak** | **Total** | **% of Total** |
|---|---|---|---|---|---|---|---|
| Negative | 8 | 5714.25 | 584 | 35078 | 470 | **50679** | **45.23** |
| Neutral | 9 | 2628.778 | 1818 | 9425 | 1315 | 39956 | 35.66 |
| Positive | 8 | 1320.25 | 980 | 3176 | 540 | 21411 | 19.11 |
| TOTAL | 25 | \ | \ | \ | \ | 112046 | 100 |

For **RQ3**, we conducted a social network analysis to better understand the patterns of interaction among X users' discussion on libraries and generative AI. This method allows us to identify key participants in the conversation, highlights how information circulates, and reveals clusters of users who frequently engage with one another. Figure 3(a) shows the mentioned network based on the top 100 most active or mentioned X users in the conversations related to libraries and generative AI. Figure 3(b) shows the top 20 user accounts and their connections.

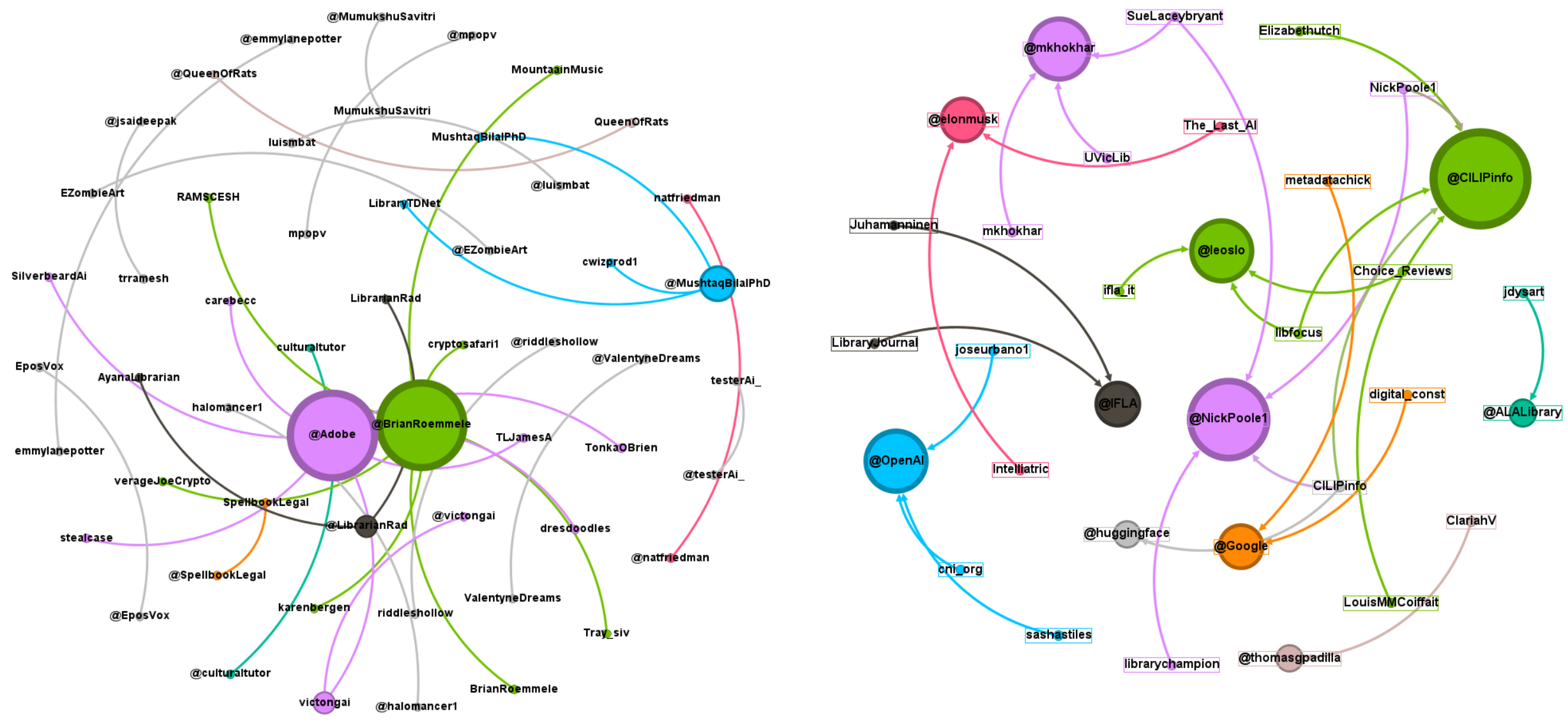

**(a)** **(b)**

**Figure 3: Social network analysis of user interactions discussing generative AI and libraries on X.** Nodes represent individual user accounts, and edges illustrate directed mention relationships. Node size corresponds to mention frequency and centrality within the network. (a) visualizes the top 100 most active or frequently mentioned users. It highlights key hubs and information brokers connecting distinct user communities. (b) presents the top 20 most influential accounts, showcasing central nodes, major influencers, and hierarchical structures in the network.

Within Figure 3(a) and Figure 3(b), each node represents an X account, while the edges (lines) indicate a direct mention from one user to another. Node size reflects the user's centrality–larger nodes are more frequently mentioned or connected within the network. Users positioned centrally are often key influencers or information brokers referring to a person or account that connects otherwise separate groups within a network. These users can help facilitate the flow of information from one group to another in the conversation. While information brokers may not always create original content, they often share, retweet, or mention information from one group and make it visible to another. Information brokers always situate between externalities/on the paths that connect different users or clusters, whereas those on the periphery are more isolated or niche contributors whose content often focuses on specific topics, perspectives, or communities. Compared to information brokers, the niche contributors in a social network often bring specialized insights or relevance within a particular area. In addition, a cluster of users suggests thematic or social communities where users frequently interact with each other, and dense clusters imply tightly connected subgroups such as librarians, tech professionals, and institutions.

Figure 3(a) shows two central hubs, *@BrianRoemmele* (highlighted in green) and *@Adobe* (highlighted in purple), with strong, diverse connections across multiple communities, suggesting these accounts are frequently mentioned by others and play key roles in information dissemination. Another hub, *@MushtaqBilalPhD* (highlighted in blue), forms a smaller but distinct network loop that tightly interacts with users like *@cwizprod1* and *@LibraryTDNet*. This might suggest topic-specific discussions or a niche representation for academic users. In addition, *@SpellbookLegal* (highlighted in orange) and *@LibrarianRad* (highlighted in gray) connect to multiple communities that are otherwise separate (e.g., *@stealcase, @victongai, @VyanaLibrarian*), indicating they are high betweenness centrality users (i.e., information brokers) that bridge groups and transfer ideas in between.

Figure 3(b), the top 20 most engaged users, shows that there are several distinct communities, with a few users acting as connectors between communities. The most influential node, *@CILIPInfo* (highlighted in green), shows many incoming connections, suggesting it is frequently mentioned by others. It acts as a hub by being connected to other central nodes, including *@NickPoole1 (highlighted in purple), @leoslo (smaller green circle)*, *Choice_Reviews, libfocus*, *digital_const,* etc. Smaller clusters, like *@NickPoole1 and @leoslo (smaller green circle),* act as a bridge between *@CILIPInfo* and other users like *@huggingface, ifla_it*, and *@libfocus*. AI-related companies, *@Google* (highlighted in orange)*, @OpenAI* (highlighted in blue), and *@huggingface* have a few strong links (Google and *@digital_const, and metadatachick; OpenAI* has strong mutual mention with *cni_org, joseurbano1, Huggingface and CILIPinfo*), indicating a small but focused group that may focus on AI tech that connects to libraries. People names *@elonmusk* (highlighted in pink) and *@mkhokhar* (highlighted in purple) stand out due to high name recognition, which seems more like broadcast nodes that get mentioned a lot but may not reciprocate themselves. In particular, *@IFLA* (highlighted in dark grey) forms a core node for international library discussions and has incoming edges from *@LibraryJournal* and *@Juhamanninen*, indicating that it may be a subject authority that is respectively mentioned, although not the most active.

Overall, the top 100 mention networks show that the users of the discussion on GenAI and libraries show strong mutual interactions within and between communities (reciprocal mentions) with high edge density in the center and tapering off into the periphery. Whereas a zoom-in observation of the top 20 presents a more hierarchical structure network with organizational accounts centered as hubs of information flow rather than individuals.

## DISCUSSION

This study explored the evolving discourse on generative artificial intelligence (GenAI) in the context of libraries, focusing on public perceptions as expressed through social media (X/Twitter). By analyzing temporal trends (RQ1), sentiment distributions (RQ2), and social network structures (RQ3), our findings shed light on how GenAI is shaping public conversations about libraries and their services.

*Temporal Trends and Thematic Shifts.* The volume of GenAI-related posts concerning libraries has shown notable fluctuations, with two top peaks in August 2023 and September 2024. These spikes happened during discussions around image generation, intellectual property/copyright, ethics, and accessibility. This temporal distribution aligns with Murayama et al. (2025), who found that GenAI-related conversations are often influenced by events such as product releases or policy debates. The high frequency of terms such as "artist," "consent," and "stock" during August 2023 suggests a strong focus on image-generation ethics, paralleling Brewer et al. (2024), who noted heightened skepticism toward AI-generated art among traditional media consumers. In contrast, the September 2024 peak reflected concerns about copyright and free access to books, possibly triggered by political campaigns or educational initiatives. The presence of educational keywords like "school" and "read" further reinforces the pedagogical relevance of GenAI. This is consistent with the research emphasis on the growing role of GenAI in instructional settings by Willet and Na (2024) and Oddone (2024). These changes highlight how public views on GenAI are shaped by technology and social factors.

*Sentiment Toward Generative AI and Libraries*. Most discussions about GenAI and libraries were negative. Over 45% of posts were negative, especially in September 2024. Neutral comments stayed steady, but positive comments were comparatively rare. The results suggest limited public optimism about GenAI in libraries. Our observation matches earlier studies (Meakin, 2024; Adegboye et al., 2024), showing that students and librarians worry about trust, misinformation, and GenAI's quality. These perceptions point to an underlying conflicted feeling: GenAI offers convenience but also creates new risks around accuracy, originality, and ethics. The relatively stable neutral sentiment may reflect a cautious curiosity or ambivalence among users as GenAI continues to evolve. Besides, positive sentiment regarding efficiency could possibly be linked to discussions about automation and enhanced productivity and user services, as identified by Harada (2025) and Semeler et al. (2024) in the context of cataloging, metadata generation, and data analysis. Future analysis could focus on exploring the connections between sentiment and discussion themes for tech-related topics and library-related topics to reveal specific areas.

*Social Network Dynamics and Influencers.* Our social network analysis revealed a structured ecosystem of interactions: both individual and institutional actors occupy influential roles in the public discussion. Central nodes such as @CILIPInfo, @Adobe, and @BrianRoemmele were frequently mentioned, indicating their prominence in shaping the discussion. These findings align with prior work by Miyazaki et al. (2024), who found that professionals in creative and technological fields were more actively engaged and positively inclined toward GenAI compared to users in education or policy domains. Notably, several users served as information brokers—accounts that connected otherwise separate communities. These included @NickPoole1, @SpellbookLegal, and @leoslo, whose positions suggest they facilitate interdisciplinary dialogue. This reflects Mannheimer et al.'s (2024) emphasis on the value of cross-sector collaboration in managing GenAI implementation. Contrary to the more peer-to-peer interaction observed with the top 100 network, the organizational accounts in the 20 network formed a more hierarchical structure. These accounts act as central hubs for broadcasting information. Such kind of dynamic contrasts with the top 100 network, which showed richer peer-to-peer interaction. Furthermore, AI companies such as @Google, @OpenAI, and @huggingface were visibly embedded within library-related discussions, signaling a growing intersection between the fields of artificial intelligence and librarianship (Chen, 2023; Cox & Tzoc, 2023). The fewer mutual mentions among these accounts and library professionals indicate room for fostering innovation in library workflows, user engagement, and information access.

## LIMITATION

While our study provides valuable insights into the public discussion on the impact of GenAI on libraries, there are several limitations that we plan to address in future work. First, we note that this study is based on data from X, which may overrepresent certain demographics and perspectives, and due to the limitation of account information, we cannot verify the connections between the posters and libraires as some of them might be professional librarians while others might just be regular users. Thus, the sample we collected could skew the generalizability of our findings. Additionally, although computational methods such as sentiment analysis are efficient and scalable, they can oversimplify the complexity of human emotions, particularly in interpreting sarcasm, irony, or context-specific language. To enhance the interpretive accuracy of our results, future work should incorporate qualitative validation, such as illustrative quotes from the dataset, to substantiate computational findings. Furthermore, the lack of direct engagement with stakeholders limits the depth of our interpretations. Future work may consider conducting interviews with relevant actors to offer richer, more grounded insights into motivations and interpretations that are not always apparent from text data alone. Finally, although our findings suggest meaningful trends, keywords, sentiment shifts, and network structures, a more explicit theoretical framework is needed to clarify how these dimensions contribute to domain-specific understanding. Anchoring our analysis within such a framework would help clarify the analytical value of each computational measure and their interplay in capturing sociocultural or behavioral dynamics.

## CONCLUSION AND FUTURE RESEARCH

This study investigates public perceptions of generative AI in the context of libraries by conducting a large-scale content, sentiment, and social network analysis of posts on X. The findings from this study highlight three overarching themes. First, the tension between promise and peril remains central to the GenAI discourse. While many users recognize the transformative potential of GenAI in enhancing efficiency, accessibility, and personalization, others raise concerns about ethical integrity, algorithmic bias, and erosion of critical thinking and engagement. Second, libraries are at a pivotal moment where they must balance innovation with a continued commitment to information ethics. In this context, professional development and digital literacy training will be essential to sustain the libraries' role as a trusted guide in the information landscape. Third, public sentiment, as expressed through platforms like X, has a critical role in shaping the social license for GenAI adoption. Influencers and organizations serve as opinion leaders whose mentions can either build trust or amplify skepticism. The observed network structure, though polarized, also reveals shared concerns and touchpoints for dialogue across communities. Our preliminary results show that the scope and intensity of the public discussion around the impact of the GenAI on libraries merit further scrutiny. Future research could build on these findings by incorporating thematic or topic modeling, cross-platform analyses, qualitative interviews, or longitudinal tracking to offer a more comprehensive understanding of how public attitudes toward GenAI and libraries continue to evolve.

## GENERATIVE AI USE

We employed GenAI specifically ChatGPT, for the following purposes: revising writing initially drafted by the authors. The output was evaluated through careful review to ensure clarity, accuracy, and alignment with academic standards. The authors assume full responsibility for the content of this submission.

## AUTHOR ATTRIBUTION

Yuan Li: conceptualization, methodology, data curation, formal analysis, writing –original draft; Teja Mandaloju: formal analysis, visualization, writing –review and editing; Haihua Chen: project administration, supervision, writing – review and editing.